\pdfoutput=1
\documentclass[
    aps,prb,twocolumn,
	groupedaddress,superscriptaddress,
	amsfonts,amssymb,amsmath,
	citeautoscript,longbibliography,
	letterpaper, nofootinbib
	]{revtex4-2}

\usepackage[utf8]{inputenc}
\usepackage[english]{babel}

\usepackage{microtype} %For better kerning and symbol-stretching
\usepackage{xspace} %For the \xspace command

\usepackage{txfonts}  %Times Roman fonts
\usepackage{txfontsb} %Addition for txfonts, including old style numerals and greek

\usepackage{bm} %Bold math symbols with \bm{} (Greek and other symbols)

\usepackage{xcolor}
\usepackage[]{graphicx} % "demo" option to disable rendering for speed
\graphicspath{{figs/}}

\usepackage[]{booktabs}
\usepackage{array}
\usepackage{layouts}
\usepackage{multirow}

\usepackage{enumerate}
\usepackage[inline]{enumitem}

\usepackage{xr}
\makeatletter
\newcommand*{\addFileDependency}[1]{% argument=file name and extension
  \typeout{(#1)}% latexmk will find this if $recorder=0 (however, in that case, it will ignore #1 if it is a .aux or .pdf file etc and it exists! if it doesn't exist, it will appear in the list of dependents regardless)
  \@addtofilelist{#1}% if you want it to appear in \listfiles, not really necessary and latexmk doesn't use this
  \IfFileExists{#1}{}{\typeout{No file #1.}}% latexmk will find this message if #1 doesn't exist (yet)
}
\makeatother

\usepackage{hyperref}
\hypersetup{colorlinks,
	linkcolor={blue!75!black!80!yellow},
	citecolor={blue!75!black!80!yellow},
	urlcolor={blue!75!black!80!yellow}
}

\usepackage[capitalize,nameinlink]{cleveref}

\crefname{subequations}{Eqs.}{Eqs.} %Specific changes to allow for Eqs.-wording when referring to a set of subequations. Label of subequations must include [subequations] as an option.
\Crefname{subequations}{Eqs.}{Eqs.}
\crefformat{subequations}{#2Eqs.~(#1)#3}
\Crefformat{subequations}{#2Eqs.~(#1)#3}
\crefname{page}{p.}{p.} %Changing from 'page' to 'p.'
\crefname{table}{Table}{Tables}
\crefname{figure}{Figure}{Figures}
\crefname{section}{Section}{Sections}

\usepackage{placeins}

\usepackage{siunitx}
\DeclareSIUnit[number-unit-product = ]\percent{\char`\%} % remove spacing for \percent

\usepackage[centering,hmargin=18mm,tmargin=29.4mm,bmargin=24mm]{geometry}

\usepackage{soul}

\usepackage{textcomp} % for \textrightarrow
\usepackage{xifthen}
\usepackage{etoolbox}
\newboolean{togglecomments}
\newboolean{toggletodos}
\newboolean{togglechanges}

\setboolean{togglecomments}{true}
\setboolean{toggletodos}{true}
\setboolean{togglechanges}{false} 

\newcommand{\textblacksquare}{$\blacksquare$}
\newcommand{\todo}[1]{\ifbool{toggletodos}%
	{\textcolor{green!60!black}{\small\textsf{{}\textsuperscript{\textsc{\textsf{todo}}}}[\ignorespaces#1]}} % if true, show comments
	{}}     % if false, do nothing
\newcommand{\comment}[2]{\ifbool{togglecomments}%
		{\textcolor{blue!70!black}{\small\sf\textsuperscript{\textsc{\textsf{\ignorespaces#1}}}[\ignorespaces#2]}} % if true, show comments
		{}}     % if false, do nothing
\newcommand{\swap}[2]{\ifbool{togglechanges}
	{\ignorespaces#2}  % revisions-only version
	{\textcolor{red!70!black}{[\ignorespaces#1]}\textrightarrow{}\textcolor{green!50!black}{[\ignorespaces#2]}}}
\newcommand{\remove}[1]{\ifbool{togglechanges}
	{}    % revisions-only version
	{\textcolor{blue}{\ignorespaces#1}}}
\newcommand{\inset}[1]{\ifbool{togglechanges}
	{\ignorespaces#1}  % revisions-only version
	{\textcolor{green!50!black}{\ignorespaces#1}}}

\newcommand{\citeremind}[1]{%
	[\textcolor{blue!75!black!80!yellow}{\textblacksquare%
		\ifthenelse{\isempty{#1}}{}{\textsuperscript{\tiny\textsf{\ignorespaces#1}}}%
	}]\xspace}

\newcommand{\appropto}{\mathrel{\vcenter{
			\offinterlineskip\halign{\hfil$##$\cr
				\propto\cr\noalign{\kern.2pt}\sim\cr\noalign{\kern-2.5pt}}}}}

\DeclareFontFamily{U}{mathx}{\hyphenchar\font45}
\DeclareFontShape{U}{mathx}{m}{n}{<5> <6> <7> <8> <9> <10>
                                  <10.95> <12> <14.4> <17.28> <20.74> <24.88>
                                  mathx10}{}
\DeclareSymbolFont{mathx}{U}{mathx}{m}{n}
\DeclareFontSubstitution{U}{mathx}{m}{n}
\makeatletter
\newcommand{\raisemath}[1]{\mathpalette{\raisem@th{#1}}}
\newcommand{\raisem@th}[3]{\raisebox{#1}{$#2#3$}}
\makeatother

\renewcommand{\paragraph}[1]{\vskip 1ex\noindent\textbf{#1.}~}

\usepackage{braket}
\usepackage[eulergreek]{sansmath}
\makeatletter
\renewcommand\@make@capt@title[2]{%
    \@ifx@empty\float@link{\@firstofone}{\expandafter\href\expandafter{\float@link}}%
    \sisetup{math-sf=\textsf}%
    \sansmath\sffamily\textbf{#1\@caption@fignum@sep}#2 % does not work with the newtx* packages unfortunately
}%

\makeatother

\usepackage{listings} % for system prompt

\setboolean{togglecomments}{true}
\setboolean{toggletodos}{true}
\setboolean{togglechanges}{false} 
\begin{document}
\title{AI-Driven Robotics for Optics}

\author{Shiekh Zia Uddin$^\bigstar$}
\email{suddin@mit.edu}
\affiliation{Department of Physics, Massachusetts Institute of Technology, Cambridge, Massachusetts 02139, USA}
\affiliation{Research Laboratory of Electronics, Massachusetts Institute of Technology, Cambridge, Massachusetts 02139, USA}

\author{Sachin Vaidya$^\bigstar$}
\email{svaidya1@mit.edu}
\affiliation{Department of Physics, Massachusetts Institute of Technology, Cambridge, Massachusetts 02139, USA}
\affiliation{Research Laboratory of Electronics, Massachusetts Institute of Technology, Cambridge, Massachusetts 02139, USA}
\affiliation{The NSF Institute for Artificial Intelligence and Fundamental Interactions}

\author{Shrish Choudhary}
\affiliation{Department of Electrical Engineering and Computer Science, Massachusetts Institute of Technology, Cambridge, Massachusetts 02139, USA}

\author{Zhuo Chen}
\affiliation{Department of Physics, Massachusetts Institute of Technology, Cambridge, Massachusetts 02139, USA}
\affiliation{The NSF Institute for Artificial Intelligence and Fundamental Interactions}
\affiliation{Institute for Data, Systems and Society, Massachusetts Institute of Technology, Cambridge, MA 02142\\
$^\bigstar$ These authors contributed equally to this work}

\author{Raafat K. Salib}
\affiliation{Department of Electrical Engineering and Computer Science, Massachusetts Institute of Technology, Cambridge, Massachusetts 02139, USA}

\author{Luke Huang}
\affiliation{Department of Electrical Engineering and Computer Science, Massachusetts Institute of Technology, Cambridge, Massachusetts 02139, USA}

\author{Dirk R. Englund}
\affiliation{Research Laboratory of Electronics, Massachusetts Institute of Technology, Cambridge, Massachusetts 02139, USA}
\affiliation{Department of Electrical Engineering and Computer Science, Massachusetts Institute of Technology, Cambridge, Massachusetts 02139, USA}

\author{Marin Solja\v{c}i\'{c}}
\affiliation{Department of Physics, Massachusetts Institute of Technology, Cambridge, Massachusetts 02139, USA}
\affiliation{Research Laboratory of Electronics, Massachusetts Institute of Technology, Cambridge, Massachusetts 02139, USA}
\affiliation{The NSF Institute for Artificial Intelligence and Fundamental Interactions}

\begin{abstract}
Optics is foundational to research in many areas of science and engineering, including nanophotonics, quantum information, materials science, biomedical imaging, and metrology. However, the design, assembly, and alignment of optical experiments remain predominantly manual, limiting throughput and reproducibility. Automating such experiments is challenging due to the strict, non-negotiable precision requirements and the diversity of optical configurations found in typical laboratories. Here, we introduce a platform that integrates generative artificial intelligence, computer vision, and robotics to automate free-space optical experiments. The platform translates user-defined goals into valid optical configurations, assembles them using a robotic arm, and performs micrometer-scale fine alignment using a robot-deployable tool. It then executes a range of automated measurements, including beam characterization, polarization mapping, and spectroscopy, with consistency surpassing that of human operators. This work demonstrates the first flexible, AI-driven automation platform for optics, offering a path towards remote operation, cloud labs, and high-throughput discovery in the optical sciences.
\end{abstract}
\maketitle 

\begin{figure}[th]
    \centering
    \includegraphics[width=\columnwidth]
    {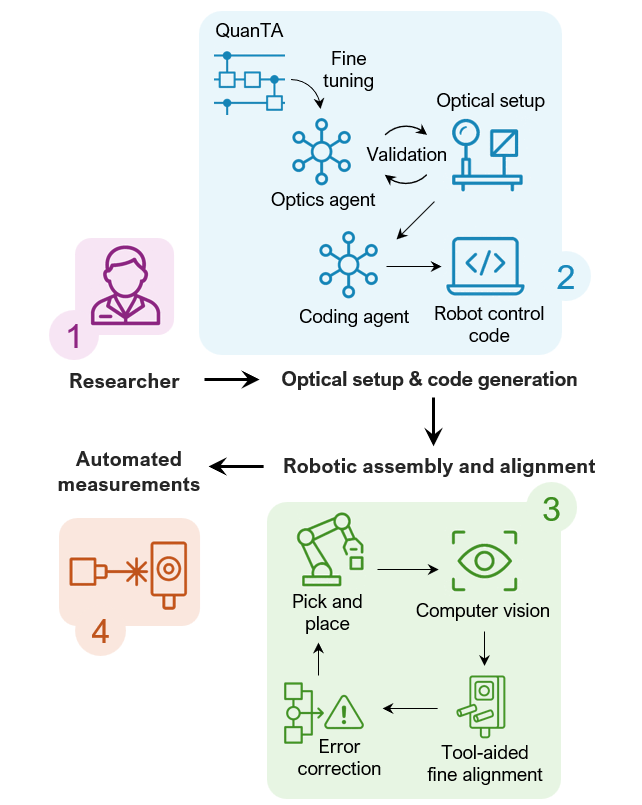}
    \caption{\textbf{Overview of the AI-driven robotics platform for free-space optics.} A researcher specifies goals for optics experiments, which are interpreted by a large language model (LLM) fine tuned using QuanTA \cite{chen2024quanta} to generate an optical setup design (``Optics agent"). The setup is then validated for accuracy and physical feasibility and translated into robot-executable code for assembly (``Coding agent"). Subsequently, a robotic arm equipped with computer vision performs optical component identification, pick-and-place operations, positional error correction, and fine alignment. Once assembled, the platform autonomously performs optical measurements such as characterization and imaging tasks.}
    \label{fig:figure1}
\end{figure}

Precise and repeatable control of experimental workflows is a central challenge across the physical sciences. In recent years, the proliferation of data‑driven methods and large‑scale artificial intelligence (AI) has been transforming how measurements are planned, executed, and interpreted~\cite{wang2023scientific, lu2024ai, krenn2022scientific}. As AI‑enabled approaches become more pervasive, the need for integrated automation in bench‑scale experiments has grown ever more urgent. Pioneering efforts in many fields have already demonstrated the utility of robotics in accelerating scientific progress. In chemistry, automated platforms now execute multi‑step syntheses and characterize reaction outcomes with minimal human intervention, enabling rapid exploration of chemical space \cite{burger2020mobile, dai2024autonomous, darvish2025organa, tom2024self, mccullough2020high, koscher2023autonomous, coley2019robotic}. Similarly, programmable fabrication systems have been harnessed to synthesize materials with targeted properties, perform manipulations with atomically precise control, and perform characterization tasks \cite{macleod2020self, mannix2022robotic, mandal2024autonomous, macleod2022self, szymanski2023autonomous}, while cloud‑based laboratories offer remote access to state-of-the-art instrumentation at large scales \cite{arnold2022cloud}.

Optics is both a fundamental science, concerned with the study of light and its interactions with matter, and an enabling technology that permeates nearly every scientific domain. It provides the basis for instrumentation in observational astronomy, advanced microscopy techniques in biology, spectroscopic and imaging tools in materials research, and precision experiments in atomic, molecular, and chemical physics. Despite successes in other disciplines, the field of optics has remained predominantly manual in its implementation. While recent studies have explored the use of AI to discover or optimize optical configurations \cite{krenn2021conceptual, krenn2025digital, arlt2024meta, landgraf2025automated}, the development of a fully autonomous, end-to-end pipeline from design to implementation remains an open problem. This inertia stems from major challenges intrinsic to free-space optical systems, which demand reconfigurability, sub-micron precision in alignment, continuous tuning of sensitive optical elements, and in-situ diagnostics that are highly susceptible to environmental fluctuations. Meeting these stringent demands requires an interdisciplinary approach: one that integrates AI, precision robotics, computer vision, engineering, and deep expertise in optical systems. An AI-driven robotic framework for optics automation could overcome these limitations, significantly accelerating experimental workflows. Such a framework would also expand accessibility to complex optical experiments and allow for the exploration of the vast space of optical configurations directly in the laboratory.

Here, we address this critical gap by introducing the first AI-driven robotics platform for free-space optics (Figure~\ref{fig:figure1}) where we aim to automate the three fundamental stages of an optical experiment: (1) design, (2) physical assembly and fine alignment, and (3) execution of measurements. At the design stage, we employ fine-tuned large language models (LLMs) that interpret user-specified goals and generate valid optical setups within the physical constraints of the laboratory environment. These LLM-generated setups are then translated into robot control code using a second LLM-based agent. A robotic arm equipped with a computer vision system performs pick-and-place operations with sub-millimeter precision, followed by fine alignment using a purpose-built robot-deployable tool. In the final stage, we demonstrate a range of automated measurements commonly performed in optics laboratories, including beam characterization, polarization mapping, transmission spectroscopy, and real-space setup optimization.

\section{Optical setup design using LLMs}
Designing free-space optical experiments requires translating high-level experimental intent into concrete spatial configurations of optical components. This requires reasoning about laboratory constraints and propagation paths. Humans accomplish this by combining domain knowledge of optical physics with spatial intuition—a capability that has remained fundamentally inaccessible to automated systems. To bridge this gap, we adopt a generative approach based on large language models (LLMs) fine-tuned to generate optical setups, enabling them to produce physically valid and robot-executable experimental designs from natural language descriptions (Figure~\ref{fig:figure1}).

To demonstrate the system’s capabilities, we focus on four canonical optical setups of varying complexity: Michelson interferometer, Mach-Zehnder interferometer, Hong-Ou-Mandel interferometer, and a 4f optical imaging system. We define an optical setup as a list of tuples, with each tuple encoding an optical component identifier, x-y position, and orientation. For simplicity, we assume all components to have a fixed z-position (height). From a single reference of each setup, we generate a dataset through geometric data augmentation using similarity transformations (translations, scaling, rotations, and reflections). These synthetic setups are subsequently filtered to ensure physical feasibility: they must lie entirely within the robot-accessible workspace, have no overlapping components, and avoid internal beam paths intersecting the robotic arm’s footprint (see Methods for details). 

For each setup in the dataset, we generate natural language descriptions using an auto-captioning step. These captions simulate user inputs with varying degrees of specificity, ranging from high-level objectives (e.g., “Give me a Mach-Zehnder interferometer”) to setups with specific requests (e.g., ``Give me a Michelson interferometer setup. The incoming laser beam direction makes an angle of 30 degrees with the x-axis and the setup should have a beam splitter positioned at approximately (20 cm, 3 cm)"). Together, these steps produce a scalable labeled dataset of physically valid, linguistically varied examples from which the model can learn to generalize, without the need for expensive optical simulation.

We then fine-tune a pretrained LLaMA3.1-8B-Instruct model for this task using Quantum-informed Tensor Adaptation (QuanTA) \cite{chen2024quanta}, a parameter-efficient fine-tuning (PEFT) method (see Methods). QuanTA utilizes quantum-inspired tensor networks to modify model weights, enabling efficient high-rank adaptations that can capture complex structural relationships while using significantly fewer trainable parameters than full fine-tuning. Our fine-tuning strategy involves two sequential stages: first, we train the model on our augmented dataset to learn the fundamental principles of generating physically valid optical layouts (structure learning); second, we continue the training with prompts containing specific user constraints/requests to enhance the model's ability to interpret and adhere to precise experimental requirements (user request understanding). Once the LLM generates an optical setup, a validation step evaluates the geometric and physical feasibility of the layout (see Methods). If the configuration fails this validation, it is returned to the model for refinement in a closed-loop correction process. Validated setups are then passed to a secondary LLM-based agent (based on Claude 3.7) which translates the symbolic layout into executable robotic movement commands for physical assembly in the laboratory (see Methods).

\begin{figure}
    \centering
    \includegraphics[width=0.9\columnwidth]
    {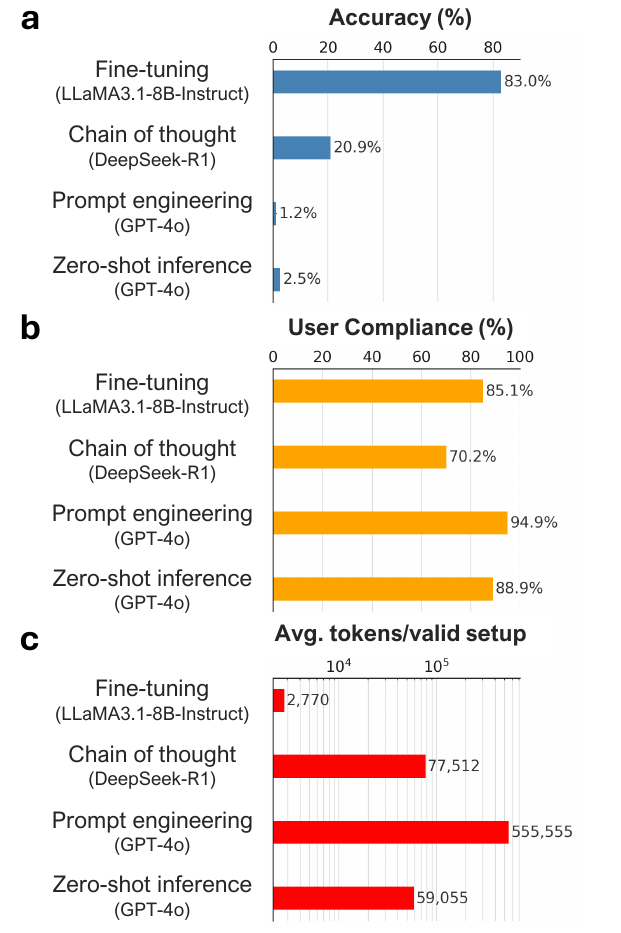}
    \caption{\textbf{Comparison of prompting strategies for autonomous optical setup generation.}
    \textbf{(a)} Accuracy across prompting strategies, averaged over all four setup types. 
    \textbf{(b)} User compliance rates, reported as the fraction of setups meeting the user-requested conditions, regardless of the validity of setups. 
    \textbf{(c)} Average token usage (log scale) per valid setup generated for each prompting strategy.}
    \label{fig:figure2}
\end{figure}

To assess the performance of our fine-tuning approach, we compare QuanTA with three alternative strategies: (1) zero-shot inference using GPT-4o, (2) prompt-engineered GPT-4o, and (3) prompt engineering combined with chain-of-thought reasoning using DeepSeek-R1 (Figure~\ref{fig:figure2}). We find that our fine-tuned model achieves the highest pre-validation accuracy ($>80\%$), defined as the fraction of correct setups over total attempts. Under user-specified constraints, such as exact placement of some components, the model maintains high accuracy while also achieving strong compliance (placement within 1 mm and 0.1$^{\circ}$ of the requested position), whereas other strategies, though compliant, fail to yield valid setups (Figure~\ref{fig:figure2}a,b). While all approaches would eventually converge to a valid setup after enough iterations through the validation step, the significantly higher accuracy of our fine-tuned model directly translates into reduced computational costs, lower token usage per valid setup, and shorter times to obtain a correct solution (Figure~\ref{fig:figure2}c).

\section{Robotic assembly and fine alignment}
Following the generation of valid and spatially aware optical setups by the LLM-based design system, the next stage involves the autonomous assembly and fine alignment of these setups using a robotic platform. This system is built around a 7-degree-of-freedom robotic arm (UFACTORY xArm7, see Methods) and is deployed in a standard optics laboratory. The robotic arm is programmed to perform high-precision placement of optical components, including mirrors, lenses, beam splitters, and detectors. We integrate the robotic arm with computer vision, LiDAR-based scanning and depth sensing, and custom-designed optical component holders and magnetic bases, all optimized for robotic manipulation. 

\begin{figure*}
    \centering
    \includegraphics[width=1.95\columnwidth]
    {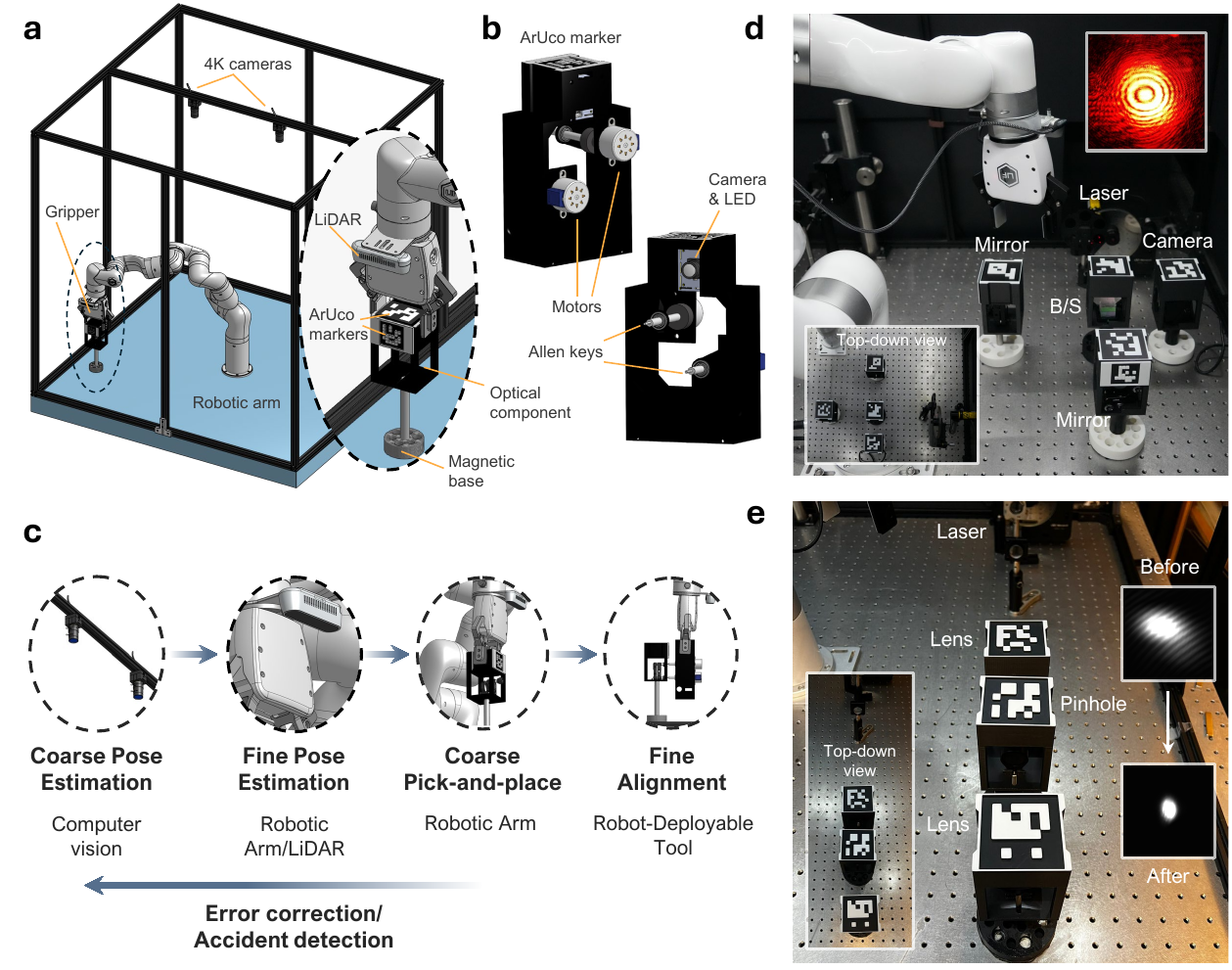}
    \caption{\textbf{Autonomous assembly and alignment of optical setups.}
    \textbf{(a)} Schematic of the robotic platform, consisting of a 7-degree-of-freedom robotic arm equipped with a gripper and a LiDAR camera, as well as external 4K cameras mounted around the workspace to provide stereo vision. Optical components are housed in custom 3D-printed casings labeled with ArUco fiducial markers for identification and positioning. The components are placed in 3D-printed magnetic bases that resist unwanted motion during handling. \textbf{(b)} Schematic of the robot-deployable fine-alignment tool, which consists of two motors with Allen keys mounted on their shafts and an onboard computer vision system enabled by a camera and a white-light LED.
    \textbf{(c)} Pipeline for autonomous assembly of optical setups. First, the computer vision system performs a coarse estimation of the position and orientation of the components in the working area using the 4K cameras. The robotic arm then refines this estimate using the onboard LiDAR sensor. Next, a coarse pick-and-place operation positions and orients the component as desired within the optical setup, achieving sub-millimeter and sub-degree precision. Finally, the robot-deployable fine-alignment tool executes micro-adjustments, achieving the sub-arcminute angular precision required for functional optical alignment. This multi-stage pipeline enables robust and repeatable assembly of complex free-space optical configurations.
    \textbf{(d)} Autonomously assembled and aligned Michelson interferometer setup in the laboratory. The insets show a top-down view of the interferometer and a camera image of the achieved interference pattern.
    \textbf{(e)} Autonomously assembled beam cleaning setup (4f optical system) in the laboratory. The insets show the beam profile before and after passing through the setup.
        }
    \label{fig:figure3}
\end{figure*}

Each optical component is placed in a 3D-printed housing that is designed to universally fit standard off-the-shelf optical components. The component is secured to a 3D-printed magnetic base that provides resistance to unwanted motion during assembly and alignment (Figure~\ref{fig:figure3}a). The housing has an ArUco fiducial marker affixed to the top surface that allows for component identification and positioning via stereo 4K visible-light cameras mounted around the workspace. These cameras perform an initial scan of the optical table, allowing the system to estimate component positions and orientations. 

Once a candidate component is identified, the robotic arm approaches it, and a LIDAR sensor mounted on the end-effector performs a high-resolution scan to refine the object's pose and geometry. This multi-modal perception pipeline ensures sub-millimeter grasping accuracy and minimizes the risk of collision or misplacement. Upon successful pickup, components are placed at their target positions and orientations according to the layout provided either by a user or by the fine-tuned LLM (see Supplementary Videos 1 and 2). During placement, an error correction step ensures both positional and angular fidelity by rechecking these parameters using the computer vision system.

For optical setups requiring precision beyond the capabilities of the initial placement, we developed a robot-deployable fine alignment tool (Figure~\ref{fig:figure3}b, also see Supplementary Video 9). This compact, motorized device is constructed to interface with the fine-adjustment knobs located on standard optical mounts, providing electronically controlled, micrometer-scale adjustments to the component’s degrees of freedom. The tool consists of an Arduino Nano microcontroller, a custom PCB with integrated motor drivers, Allen keys mounted on motor shafts, and a dedicated camera and LED illumination system. For fine alignment tasks, the robotic arm autonomously picks up the fine alignment tool and inserts the Allen keys of the tool into the adjustment knobs of optical components. Alignment adjustments are performed iteratively, guided by feedback from cameras or photodetectors previously placed along the beam path. The complete robotic pipeline is shown schematically in Figure~\ref{fig:figure3}c (see Methods for more details on the computer vision system, robot pick-and-place operations, custom housing and bases, construction of fine-alignment tool, and correction of positional errors).

To demonstrate the capabilities of the robotic assembly and fine-alignment system, we first ask the AI-driven robotic system to autonomously assemble and align a Michelson interferometer from scratch (Figure~\ref{fig:figure3}d, Supplementary Videos 3 and 4). The optical layout for the interferometer is generated by the fine-tuned LLM in response to a user-specified constraint on the incoming laser beam direction. Based on the generated design, the computer vision system identifies the required components — two mirrors, a beam splitter, and a camera — and the robotic arm subsequently places them on the optical table (Figure~\ref{fig:figure3}d). Following initial assembly, the robotic arm deploys the fine-alignment tool to adjust the orientation of the mirrors, ensuring the precise overlap of the two beams of the interferometer. Feedback for the alignment is obtained by blocking each arm and centering the beams in the camera's field of view. The final interferometer and the resulting stable interference pattern are shown in Figure~\ref{fig:figure3}d.

As a second demonstration, we ask the AI-driven robotic system to autonomously assemble a 4f beam cleaning setup consisting of two lenses and a pinhole. As before, the optical layout is generated by the fine-tuned LLM after specifying the incoming laser direction. The robotic arm places the components to create a 4f setup, and a camera images the beam before and after setup assembly (Figure~\ref{fig:figure3}e, also see Supplementary Video 10). 

\section{Automated Optical Measurements}

\begin{figure*}
    \centering
    \includegraphics[width=2\columnwidth]
    {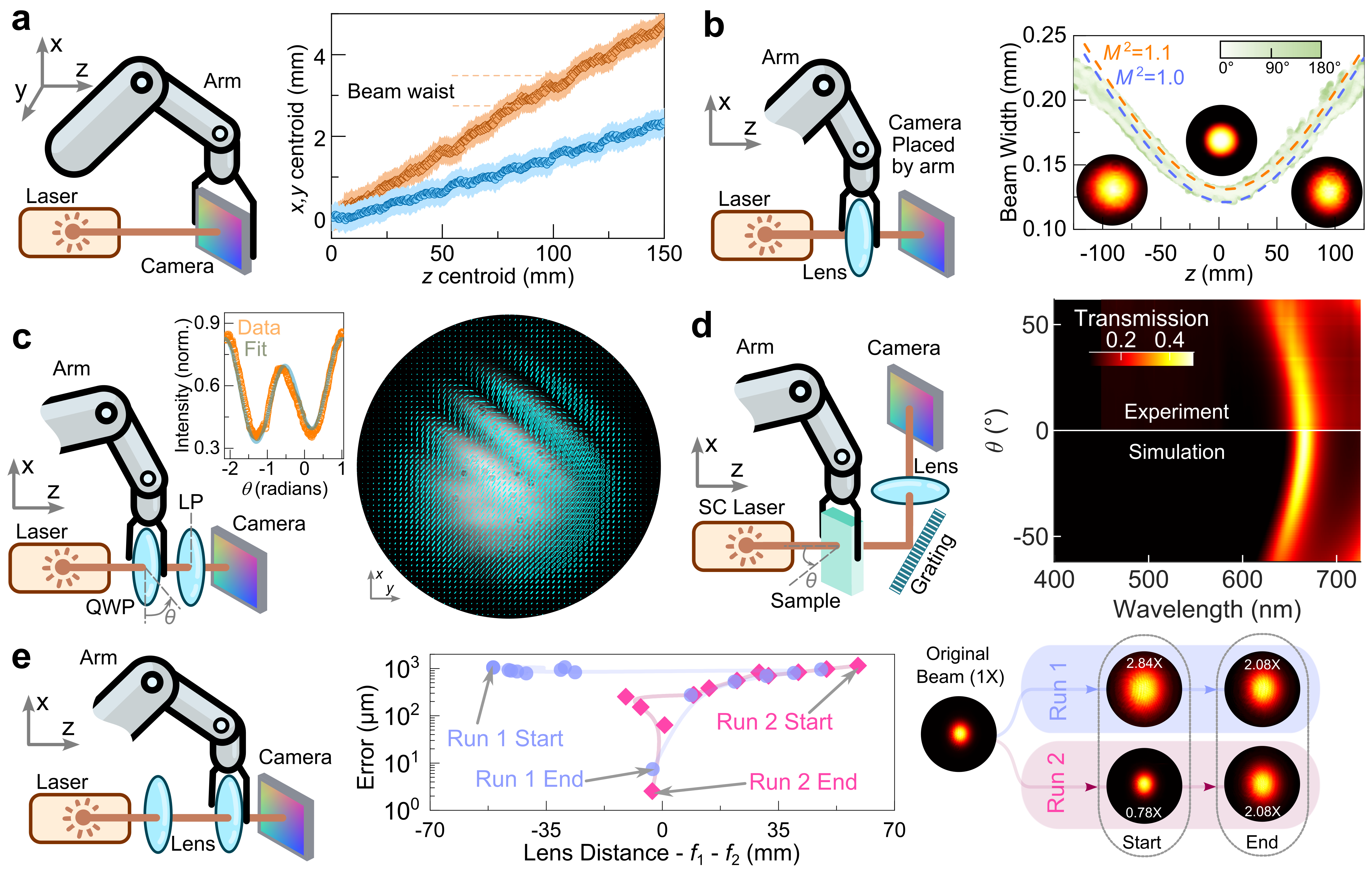}
    \caption{\textbf{Automated optical measurements, characterization and optimization tasks.} 
    \textbf{(a)} The robotic arm linearly moves a camera, tracking the centroid of the laser beam to measure its direction and width. In a typical configuration, beam angles along the x, y, and z axes are measured to be 88.16$^\circ \pm$ 0.05$^\circ$, 89.16$^\circ \pm$ 0.05$^\circ$, and 2.03$^\circ \pm$ 0.04$^\circ$, respectively, averaged over 230 measurements 
    \textbf{(b)} The robotic arm places a lens in the beam path and moves the camera in the z-direction, capturing images. These data are used to calculate the beam matrix.
     \textbf{(c)} The robotic arm assembles a setup consisting of a camera and a linear polarizer (LP) in the beam path. The beam passes through a birefringent film (BRF), creating spatial variations in the polarization. The robotic arm holds a quarter-wave plate (QWP) and rotates it, while the camera captures images. These images are used to calculate the Stokes parameters and plot polarization ellipses for every point in space, superimposed on the intensity image. 
    \textbf{(d)} The robotic arm assembles a setup consisting of a grating and a camera to create a spectrometer and then holds and rotates a multilayer photonic crystal to obtain the angle- and wavelength-resolved transmission spectrum.
    \textbf{(e)} The robotic arm places a $f_1=60$ mm focal length convex lens in the beam path and repeatedly places a $f_2=125$ mm focal length lens at different locations with the goal of achieving $f_2/f_1=2.083\times$ beam expansion. 
 Two such runs with different initial positions of the lens are shown; both ultimately converge to the same lens distance. Beams before and after real-space optimization are also shown.
        }
    \label{fig:figure4}
\end{figure*}

Following the successful autonomous assembly and fine alignment of optical setups, we proceed to execute automated optical measurements with our platform. While the assembly process involves static placement of components, many optical experiments require dynamic operations such as translating, rotating, or adjusting optical elements in real time. These dynamic measurement tasks demand the same level of precision, robustness, and autonomy as the assembly process, but introduce additional challenges related to motion control. Moreover, measurements are also integral to the assembly phase itself—for instance, verifying laser alignment and characterizing beam quality during setup construction. Thus, the integration of real-time optical diagnostics and measurement capabilities is critical to achieving fully autonomous experimentation.

Leveraging the robotic platform’s ability to manipulate optical elements with high spatial and angular accuracy, we demonstrate automated execution of optical measurements. In this section, we show this functionality across five examples of distinct tasks: (1) beam direction and waist tracking, (2) beam quality characterization through beam matrix measurement, (3) polarization mapping via spatially resolved Stokes parameter extraction, (4) angle- and wavelength-resolved spectroscopy of multilayer photonic crystals, and (5) real-space optimization of a beam expansion setup.

To determine the propagation direction of a collimated laser beam, the robotic arm picks up a camera module and positions it such that the beam is centered within the field of view. The robotic arm then translates the camera while keeping the beam centered in the camera's field of view. By recording the beam centroid through this motion, the beam's angle relative to the laboratory axes is calculated using a linear regression fit (see Supplementary Video 5). Similarly, the beam waist can be directly measured from the camera images (Figure~\ref{fig:figure4}a).

Precise spatial characterization of laser beams is essential in a wide range of experiments, from beam shaping and mode matching to the calibration of complex optical systems. The complete spatial characterization of the beam can be determined via a measurement of the second-order moments of the Wigner distribution, commonly referred to as the beam matrix, which represent the beam's spatial intensity and phase in both position and momentum (or angle) space. This matrix records beam properties such as width, divergence, astigmatism, and ellipticity. The components of the beam matrix can be measured using cylindrical and spherical lenses and a camera to image the beam at different planes along the propagation axis and at different positions.

To measure the beam matrix, the robotic arm places down a camera and translates a lens in the beam path, capturing the transverse intensity profile at multiple propagation distances (see Supplementary Video 6). From these images, second-order spatial and angular moments are computed, yielding a full $4 \times 4$ beam matrix (Figure~\ref{fig:figure4}b):

\begin{align*}
M = \begin{pmatrix}
\langle x^2 \rangle & \langle xy \rangle & \langle x\theta_x \rangle & \langle x\theta_y \rangle \\
\langle xy \rangle & \langle y^2 \rangle & \langle y\theta_x \rangle & \langle y\theta_y \rangle \\
\langle \theta_x x \rangle & \langle \theta_x y \rangle & \langle \theta_x^2 \rangle & \langle \theta_x \theta_y \rangle \\
\langle \theta_y x \rangle & \langle \theta_y y \rangle & \langle \theta_x \theta_y \rangle & \langle \theta_y^2 \rangle
\end{pmatrix}
&= \\[1ex]
10^{-7} \begin{pmatrix}
0.13 \;\mbox{m}^2 & 0.06 \;\mbox{m}^2 & 0.08\;\mbox{m}& -0.02 \;\mbox{m}\\
0.06 \;\mbox{m}^2 & 0.12 \;\mbox{m}^2 & 0.27 \;\mbox{m}& 0.00 \;\mbox{m}\\
0.08 \;\mbox{m} & 0.27 \;\mbox{m} & 2.80 & 0.04 \\
-0.02 \;\mbox{m} & 0.00 \;\mbox{m} & 0.04 & 3.06
\end{pmatrix}
\end{align*}

Here, $x$ and $y$ are the spatial coordinates of the beam, and $\theta_x$ and $\theta_y$ are the angular divergences in the $x$ and $y$ directions, respectively. The diagonal elements describe the beam's spread (e.g., $\langle x^2 \rangle$ is the beam width in the $x$-direction), while the off-diagonal elements describe correlations between position and angle or between different spatial directions. The effective beam propagation ratio is measured to be $M_\text{eff}^2 = \frac{4\pi}{\lambda}[\det M]^{1/4}=$ 1.07$\pm$0.03, where $\lambda = 632.8$ nm is the wavelength of light.

We next turn our attention to the characterization of spatially varying polarization fields, which is essential for identifying polarization singularities and vectorial field distributions in beam optics applications. To show this functionality, the robotic arm assembles a setup consisting of a linear polarizer and camera and then inserts and rotates a quarter-wave plate (QWP) along the beam path (see Supplementary Video 7). The intensity at each camera pixel $I(\theta)$ as a function of QWP angle $\theta$ follows:

\[
I(\theta) = \frac{1}{2} \left( S_0 + S_1 \cos^2 2\theta + S_2 \sin 2\theta \cos 2\theta + S_3 \sin 2\theta \right)
\]

By fitting this expression at each pixel, the spatially resolved Stokes parameters $S_0, S_1, S_2, S_3$ are extracted. From these, the degree of polarization and the orientation of polarization ellipses are computed and superimposed on the beam intensity image (Figure~\ref{fig:figure4}c).

We next demonstrate automated angle-resolved spectral characterization of multilayer photonic crystals, a task commonly performed to measure the bands of periodic photonic materials. This measurement typically requires precise and repeatable angular sweeps, making it labor-intensive and sensitive to alignment errors when performed manually. The robotic arm first assembles a setup consisting of a diffraction grating and a camera positioned along the path of a supercontinuum (SC) laser beam. It then picks up a multilayer photonic crystal (consisting of alternating layers of Si and SiO$_2$ with an embedded defect layer to form a cavity) housed in a 3D-printed casing. The arm then rotates the photonic crystal through a range of incidence angles (see Supplementary Video 8). At each angle, the transmission spectrum is recorded, yielding the complete wavelength- and angle-resolved transmission spectrum (Figure~\ref{fig:figure4}d) (see Methods for more details).

To demonstrate the robotic arm's capability for real-space optimization, we construct a beam expander consisting of two convex lenses with focal lengths $f_1 = 60$ mm and $f_2 = 125$ mm. While the theoretical optimal separation is $f_1 + f_2$, practical considerations such as lens positioning within housing assemblies and finite lens thickness necessitate experimental optimization to achieve the desired beam expansion ratio of $f_2/f_1$.  To demonstrate this real-space optimization, the robotic arm first positions the first lens. An adaptive gradient descent algorithm then determines the optimal placement of the second lens by iteratively varying its coordinates and minimizing the error function defined as the absolute difference between the target and measured beam widths. Figure \ref{fig:figure4}e shows the optimization error as a function of lens separation for two experimental runs, each initialized from positions displaced to either side of the theoretical optimum. Both optimization sequences converge to the correct lens position roughly within ten iterations, successfully achieving the target beam expansion ratio of $f_2/f_1=2.08$, as confirmed by the beam profile measurements shown for the original beam and the beams recorded before and after optimization for each run.

These five demonstrations collectively highlight the flexibility of robotic systems for executing high-precision optical measurements. Together, they establish that a robotic arm can serve as a versatile experimental tool in optics and photonics, effectively replacing a range of custom, purpose-built characterization instruments with a single reconfigurable platform.

\section{Discussion and conclusion}
We have presented the first AI-driven robotic platform capable of designing, assembling, aligning, and operating free-space optical experiments, demonstrating that full-stack automation in optics is both accurate and technically achievable. By integrating artificial intelligence, a robotic arm guided by computer vision, and modular custom alignment tools, we have established an automation pipeline that substantially reduces manual intervention and unlocks a new regime of speed, reproducibility, and remote operability in optics.

For LLM-based optical setup generation, we have shown that our fine-tuning approach significantly outperforms conventional prompt-engineering-based or zero-shot methods with respect to physical validity, user compliance, and token-efficiency. This indicates that domain-specific fine-tuning, combined with structural constraints, can produce models capable of reasoning over physically grounded design spaces—a capability of increasing importance for AI-assisted experimentation across disciplines.

The robotic system addresses a persistent bottleneck in optics: the precise placement and alignment of free-space optical components. Our demonstration of automated optical measurements—including beam characterization, polarization mapping, spectroscopy, and optimization tasks—illustrates the breadth of tasks amenable to full automation. These examples represent generalizable workflows that can be extended to material characterization, imaging applications, and remote beam monitoring, particularly in hazardous environments such as high-power laser and X-ray facilities.

Looking ahead, several opportunities exist to further enhance this platform. One promising direction is the incorporation of reinforcement learning algorithms for setup alignment and continuous optimization under real-world non-ideal conditions~\cite{kober2013reinforcement, ghavamzadeh2015bayesian}. This is particularly relevant in optics, where the reward signal can be sparse and highly intermittent during alignment, posing a challenge that may be overcome with modern machine learning techniques. Another avenue is the integration of simulation-in-the-loop frameworks, where optical modeling tools inform both LLM-based setup generation and robotic execution, leading to self-optimizing workflows. Finally, remote operation and cloud-based optical laboratories will enable distributed and democratized access to advanced optical experimentation.

\section{Acknowledgments}
We acknowledge support from the National Science
Foundation under Cooperative Agreement PHY-2019786
(The NSF AI Institute for Artificial Intelligence and Fundamental Interactions). S.C., R.K.S. and L.H. acknowledge funding from the MIT Undergraduate Research Opportunities Program (UROP). Z.C. acknowledges support from the Mathworks Fellowship. This work is also supported in part by the U. S. Army Research Office through the Institute for Soldier Nanotechnologies at MIT, under Collaborative Agreement Number W911NF-23-2-0121. The computations in this paper were partly run on the FASRC Cannon cluster supported by the FAS Division of Science Research Computing Group at Harvard University. We acknowledge funding from MIT Generative AI Impact Consortium (MGAIC). We also acknowledge support from Parviz Tayebati. Copyediting of this manuscript was performed in part using ChatGPT models (versions 4o and 5). We also thank Dr. Paola Rebusco for critical reading and editing of the manuscript.

\section{Author contributions}
M.S., S.V. and S.Z.U. conceived the original idea with input from D.R.E.; S.Z.U., S.V., S.C., and R.K.S. developed the robotic platform; S.Z.U., S.C. and R.K.S. performed all experiments and measurements with input from S.V. and L.H.; Z.C. and S.V. generated the optical setup dataset, performed fine tuning of the LLMs and validation of generated setups. The manuscript was written by S.V. and S.Z.U. with inputs from all authors. M.S. supervised the project.

\section{Data Availability}
The codes that support the plots within this paper and other findings of this work are available online in~\cite{OptSetGen, LLM-eval, AUTO-Arm, AprilTag-Detection, CAD-Engraver}. Correspondence and requests for materials should be addressed to S.Z.U. (suddin@mit.edu) and S.V. (svaidya1@mit.edu).

\section{Conflict of interest}
S.Z.U., S.V., S.C., R.K.S., L.H. and M.S. are seeking patent protection for some ideas in this work (U.S. provisional patent application no. 63/745,115). The remaining authors declare no conflicts of interest.

\section{Supplementary Material: Videos}
The following videos show key capabilities of the robotic system; click the embedded links to view them.
\begin{itemize}
    \item \href{https://youtu.be/aJdHo2FGdSs}{Video 1}: Pick-and-place of optical components
    \item \href{https://youtu.be/zfZY_MpxlpI}{Video 2}: Repeated and consistent pick-and-place of components
    \item \href{https://youtu.be/bUaSw-KOAX8}{Video 3}: Assembly and teardown of LLM-generated Michelson interferometers.
    \item \href{https://www.youtube.com/watch?v=fxjXQAmyJEk}{Video 4}: Complete assembly and fine alignment of a Michelson interferometer.
    \item \href{https://youtu.be/nc8Sr8fSde4}{Video 5}: Measurement of beam direction and width.
    \item \href{https://youtu.be/8VmDV0cLOeI}{Video 6}: Measurement of beam $M^2$.
    \item \href{https://youtu.be/jFPWFat6xis}{Video 7}: Polarization mapping measurement.
    \item \href{https://youtu.be/_PQmDLJ37CA}{Video 8}: Measurement of angle-resolved transmission spectrum of a photonic crystal.
    \item \href{https://youtu.be/ocdLUdCUIfI}{Video 9}: CAD model of the robot-deployable fine-alignment tool.
    \item \href{https://youtu.be/P7FZhsYfmic}{Video 10}: Building a beam cleaning 4f system.
    \item \href{https://youtu.be/3B9kBU7TpOY}{Video 11}: Real-space optimization to create a $2\times$ beam expander.
\end{itemize}

%\section{Supplementary Material: Images}
\setcounter{figure}{0}
\renewcommand{\thefigure}{S\arabic{figure}}

\section{Methods}
\subsection{Optical setup dataset generation}
To construct a diverse and laboratory-aware dataset of optical layouts, we developed a data augmentation pipeline centered around a library of canonical optical setups: Michelson interferometer, Mach-Zehnder interferometer, Hong-Ou-Mandel interferometer, and a 4f optical system. Each setup is encoded as a list of component tuples—each specifying a component ID (e.g., mirror, beam splitter), spatial coordinates (x, y in cm), and optical axis orientation (in degrees). We assumed all components to have a fixed height in the z-direction. The first tuple with the component ID ``000" is always a ray representing the incoming laser beam. The spatial coordinates within this tuple are those of any point lying on the ray and the orientation component is its angle with the x axis.

To expand this initial set and introduce geometric variability, we applied randomly chosen similarity transformations to each reference layout. Specifically, for each predefined reference setup, we generated multiple setups via random combinations of scaling, in-plane rotation, translations, and mirroring across the x- and/or y-axis. Each transformation was applied to the full component list, updating both spatial coordinates and optical axis orientations accordingly.

Resulting setups were subjected to rigorous physical validation to ensure experimental feasibility. Validation criteria included: (1) all components must lie within the reachable circular workspace of a 61 cm radius; (2) no component may intersect the robot’s 15 cm radius circular footprint at the origin; (3) inter-component spacing must exceed individual component dimensions (7.5 cm × 6.4 cm) to avoid spatial overlap; and (4) internal beam paths must not intersect the robotic arm’s footprint, determined by constructing a complete graph from the component coordinates and checking for overlap. Setups failing any of these constraints were discarded. The code for optical setup generation and validation is available in~\cite{OptSetGen}.

\subsection{LLM fine tuning}
To enable the translation of natural language experimental goals into precise and structured optical layouts, we fine-tuned the LLaMA3.1-8B-Instruct model using a parameter-efficient fine-tuning (PEFT) strategy. Specifically, we employed Quantum-informed Tensor Adaptation (QuanTA) \cite{chen2024quanta}, a method inspired by quantum circuit structures and tensor networks that enables efficient high-rank updates to pre-trained model weights. Compared to conventional low-rank adaptation methods, QuanTA offers increased representational capacity with comparable computational efficiency, making it well-suited for capturing the complex structural and geometric constraints inherent to optical setup design.

A consistent system prompt was employed throughout both the fine-tuning and inference stages to clearly define the task scope, specify the desired output format (structured lists of component tuples: ID, x, y, orientation), and impose fundamental constraints such as workspace boundaries and component non-overlap. Full details of the system prompt and dataset specifications are provided in~\cite{LLM-eval}.

\subsubsection{Fine tuning protocol}

\begin{figure*}
    \centering
    \includegraphics[width=2\columnwidth]
    {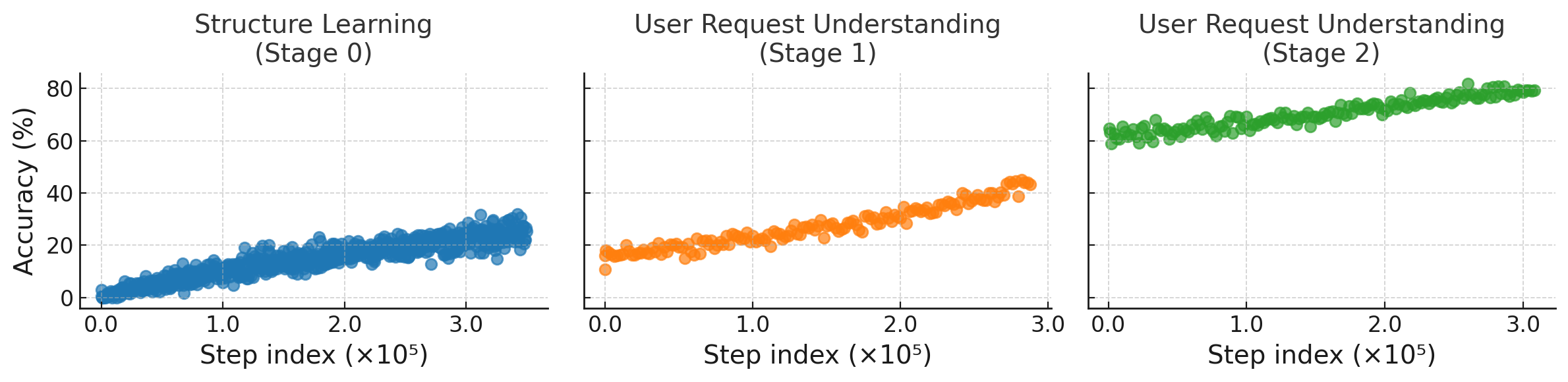}
    \caption{\textbf{Training curves for LLM accuracy.}  Accuracy (\%), defined as the fraction of correct setups over total attempts, of the QuanTA fine-tuned LLaMA model as a function of training step index (scaled in units of \(10^{5}\)). Each point corresponds to an evaluation checkpoint. The three stages are: structure learning (Stage 0), user request understanding with three requests (Stage 1), user request understanding with random requests (Stage 2).}
    \label{fig:figureS1}
\end{figure*}

Our fine-tuning protocol involved two sequential stages: 
\emph{Structure Learning} -- In the first stage, the LLaMA3.1-8B-Instruct model, initialized with the system prompt, was fine-tuned on the optical setup dataset (described above) comprising an equal but random mixture of pairs of auto-captioned natural language descriptors and their corresponding structured optical layouts for the four photonic setups. The primary goal of this stage was to enable the model to learn and internalize the knowledge behind the production of syntactically correct and physically plausible component arrangements, consistent with the basic spatial rules defined during dataset generation.

\emph{User Request Understanding} -- Following the initial structure learning, the model underwent further fine-tuning using prompts representing more specific user requests, such as prescribed component positions or beam path characteristics (e.g., ``Describe a compact Michelson interferometer optical system. The system should have a laser beam entering at 318.9 degrees."). While the target outputs remained structured component lists, the prompts introduced more nuanced constraints, improving the model's ability to flexibly interpret and satisfy user-defined experimental requirements. This stage is further split into two sessions. In the first session, we train on samples where the user always asks for a fixed number of three requests, allowing the model to understand user requests in a relatively simple format. We relax this condition in the second session, to further guide the model in interpreting flexible user requests.

To further improve the diversity of samples, parts of user prompts were generated independently using various templates, with varying precision requirements, language style, notation differences, etc. Together with the randomly generated setup coordinates and angles, this created an exponentially large data distribution, accurately mimicking possible real human inputs, and preventing the model from relying on specific patterns in the generated sample pairs (examples of input output pairs can be found in the~\cite{LLM-eval}).

The training curves across all three stages are shown in Figure~\ref{fig:figureS1}.

\subsubsection{Hyperparameters}
During fine-tuning, we chose QuanTA dimensions of 64-16-4, resulting in 143.13M training parameters (1.75\% of the full model). We used the AdamW optimizer with a weight decay of 0.01 throughout the training. During the first stage, and the first session of the second stage, we used a peak learning rate of 0.00001, and during the second session of the second stage, we chose a peak learning rate of 0.000002. For each stage or session, we started the training with an initial linear warm-up, followed by a cosine decay. The first stage was trained for 349500 iterations. The first session of the second stage was trained for 288000 iterations, and the second session of the second stage was trained for 308000 iterations, with the checkpoint at 282000 iteration picked based on accuracy evaluation on a validation dataset independently generated from both the training set and the test set. We used a batch size of 8 throughout the training, and each stage or session was trained for only a single epoch, with no samples seen twice by the model. The full training required $\sim$8500 GPU hours with A100 GPUs and 80GB VRAM on each GPU.

\subsubsection{Other prompting strategies}
To establish baseline performance benchmarks, we evaluated two contemporary large language models: GPT-4o (May 13, 2024 version) and DeepSeek-R1 (Jan 20, 2025 version). For the GPT-4o model, we assessed performance under two distinct prompting conditions: (1) zero-shot prompting, providing the model solely with the user request and the required output format specification, with minimal examples for demonstrating the output format (the same prompt as the one used during fine-tuning of the LLaMA3.1-8B-Instruct model) and (2) few-shot prompting, augmenting the prompt with illustrative examples of valid pairs of user input and model output. For the DeepSeek-R1 model, we employed the same few-shot prompt, while allowing the model to perform a few-shot Chain-of-Thought (CoT) reasoning. This approach was structured to guide the model through step-by-step reasoning for optical layout construction, supported by representative examples. 

\subsection{Validation of optical setups and LLM performance}
To ensure that optical setups generated by the LLM were accurate and adhered to physical and geometric constraints, we employed a validation procedure that extended the checks used during optical dataset generation. First, we applied a set of rule-based spatial constraints to the model-generated setup: all components must lie within the robot’s 61 cm radius reach, remain outside the 15 cm radius circular footprint of the robot at the origin, maintain non-overlapping placement based on their 7.5 cm × 6.4 cm physical dimensions, and internal beam paths must not intersect the robotic arm’s footprint. 

Passing these spatial checks does not guarantee that the laser beam will correctly propagate through the setup, as the component orientations or positions may still be physically inaccurate. To address this, we performed an additional validation step that verified the structural correctness of the generated setup against a known reference. We did this by attempting to identify a similarity transformation that mapped the LLM-generated setup onto one of the predefined reference setups of the same type. This is framed as a geometric alignment task, where the candidate transformation must preserve component identities and maintain approximate spatial relationships within tolerance bounds. If such a transformation exists, the generated setup is deemed valid due to its equivalence to the corresponding reference configuration. This step ensured that the model had not only obeyed spatial constraints but also produced a layout that structurally corresponds to an accurate optical arrangement of components. 

The accuracy of all prompting strategies was measured against a newly generated test dataset with 128 samples for each setup, utilizing a generation temperature of $0.5$ to moderate output randomness. The accuracies and user compliance scores are reported in Figure~\ref{fig:figure2} of the main text.

\subsection{Generation of robot motion code}
In this part of our workflow, we utilized Claude 3.7, an LLM, to transform high-level interferometer design specifications into executable robotic control code. By leveraging pre-defined wrapper functions that discretize the optical assembly task, we systematically translated the design parameters directly into robotic instructions. The process involved providing Claude 3.7 with a list of tuples with component specifications in the format $[\mbox{component ID}, \mbox{x position}, \mbox{y position}, \mbox{orientation angle}]$ from the LLM design along with a mapping between component identifiers and their corresponding ArUco markers of the hardware for computer vision recognition. After describing the system's functional capabilities through four wrapper functions (system initialization, component pickup via visual recognition, precise placement, and robotic arm homing), Claude generated a Python main function that systematically processed the interferometer design. The resulting code initialized the vision and robotic systems, parsed the design data, appropriately filtered non-robotic components, and implemented an assembly sequence that handled component identification, retrieval, precise placement, and safe arm positioning between operations. This procedure achieved the optical geometry specified by the LLM optical design.

\subsection{Robotic arm specifications}

\begin{figure}
    \centering
    \includegraphics[width=0.66\columnwidth]
    {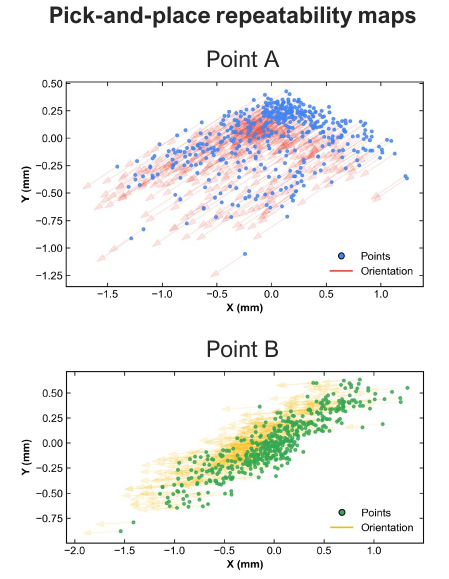}
    \caption{\textbf{Pick-and-place repeatability maps.} The robot was programmed to repeatedly transfer an object between two target points (A and B), separated by 5 cm, at a fixed orientation for each point. The scatter of points indicates the final placement positions, while arrows denote the measured orientations at each trial.}
    \label{fig:figureS2}
\end{figure}

The robotic arm is a UFACTORY xArm 7 with 7 degrees of freedom, a payload capacity of 3.5 kg, and a reach radius of 61 cm. The manufacturer reports a nominal repeatability of $\pm$0.1 mm; however, under our experimental conditions and measurement pipeline, we determined practical spatial and angular placement precisions of $\pm$0.5 mm and $\pm$0.2$^{\circ}$, respectively. These values were obtained by repeatedly commanding the robot to transfer an object between two fixed locations separated by 5 cm, and extracting the resulting distributions of object positions and orientations (Figure~\ref{fig:figureS2}). The observed scatter plots, which we refer to as pick-and-place repeatability maps, highlight the robot’s capacity for consistent handling and placement of objects.

\subsection{Computer vision system}
The robotic system’s vision pipeline is built around a stereo vision setup composed of two 4K ELP cameras mounted 23 cm apart. Initial calibration was conducted using a standard checkerboard pattern, with images collected at varying positions on the optical workbench. Intrinsic and extrinsic parameters of the stereo pair were obtained using the OpenCV cv2.calibrateCamera and cv2.stereoCalibrate functions. The calibration outputs were validated by comparing the predicted baseline offset to the physical camera separation, confirming sub-millimeter discrepancies. These calibration parameters were stored in a serialized NumPy archive (stereo\_calibration2.npz) for use in subsequent 3D position estimations during coarse alignment operations.

To enable robust and accurate object detection, we employed the ArUco fiducial markers. Each optical component is tagged with a unique identifier drawn from a 6×6 binary dictionary comprising over 800 distinguishable codes (see~\cite{CAD-Engraver} for code). These tags are detected using the cv2.aruco module, which computes the component's pose via the cv2.solvePnP algorithm using previously derived camera intrinsics. While the x and y positional estimates from ArUco detection were sufficiently accurate for coarse positioning, the z-coordinate was instead inferred from a depth measurement. For this, we deployed a robot-mounted Intel RealSense depth camera equipped with LiDAR. After the ArUco tag was centered in the camera's field of view, the LiDAR sensor provided a high-accuracy depth estimate, enabling precise z-coordinate acquisition. The robot’s end-effector was then translated vertically from a known camera-to-gripper offset to achieve accurate engagement with the object.

The precision and robustness of this system allowed for successful detection and manipulation even with smaller ArUco tags, as demonstrated in the fine alignment tool deployment. To facilitate consistent image acquisition, the workspace was illuminated with diffuse white-light LED strips integrated into the overhead structure. Additional directional LEDs were positioned in the fine-alignment tool to enhance tag visibility. To suppress glare and reduce errors from ambient light sources, light-blocking panels were installed around the frame perimeter.

To support micro-scale adjustments, we developed an auxiliary vision routine dedicated to centering the fine-alignment tool. An ArUco marker affixed to the housing of the tool is tracked to guide lateral adjustments, aligning the tool precisely with rotational knobs or alignment targets. This system has been validated across a range of component geometries and operates reliably under varied lighting conditions.

A potential limitation of the current implementation is the need for uniquely assigned ArUco tags across all components. In practice, this is mitigated by the large dictionary size and spatial separation of simultaneously visible tags. In multi-robot configurations, each robot can utilize a dedicated camera and ArUco ID space, supporting scalability for distributed collaboration in large or reconfigurable optical assemblies. The code for computer vision is available in~\cite{AUTO-Arm, AprilTag-Detection}

\subsection{Optical component casing and bases}
To facilitate reliable robotic handling of diverse optical components, we designed and fabricated universal housings and bases using fused deposition modeling (FDM) 3D printing with polylactic acid (PLA) filament using  the Bambu X1E 3D printer. Each housing was tailored to securely enclose optical components, while presenting a standardized external geometry. This uniformity eliminates the need for the robotic arm to adapt dynamically to variations in component shapes and sizes, simplifying grasping and manipulation.

An indentation feature was incorporated into each housing, precisely aligned to the robot’s claw gripper dimensions. This design ensures repeatable, stable engagement of the gripper during pickup and transport, significantly reducing the likelihood of slippage or misalignment during handling.

For optical component positioning and identification, ArUco fiducial markers were affixed to the top and side surfaces of each housing. Using these fiducial markers, the system achieves robust visual tracking, even under moderate occlusion or variable lighting conditions.

To ensure stable and repeatable placement of optical components, each unit is supported by a custom-designed base optimized for tabletop experimental environments. The bases were fabricated using fused deposition modeling (FDM) 3D printing with polylactic acid (PLA) filament. To prevent sliding and enhance frictional stability, each base was equipped with a rubberized bottom layer. Additionally, multiple neodymium magnets were embedded within the lower section of the base. The combined effect of the rubber and magnets significantly increases the frictional force between the base and the optical table, reducing unwanted sliding, tipping, and rotation under typical laboratory vibrations or minor external disturbances (e.g., those caused by the insertion of the fine-alignment tool).

\subsection{Robotic pick-and-place operations}
The robotic arm executes pick-and-place operations by integrating global and local visual feedback with precise end-effector control. For each task, the system is first provided with the unique identifier corresponding to the desired optical component. The global stereo vision system, composed of two overhead 4K cameras, detects the ArUco markers affixed to the component's casing and computes its $(x, y)$ position and orientation relative to the optical table.

Following coarse localization, the robotic arm approaches the component using a pre-defined trajectory that first elevates the end effector above the tallest components in the workspace, thereby minimizing collision risks before descending to engage the target. During the final approach phase, a secondary camera mounted on the robot’s end-effector provides high-resolution, close-range visual feedback. This enables fine alignment of the gripper with the standardized indentation designed into each casing, ensuring a secure and precise engagement during pickup.

Upon successful acquisition, the robotic system can perform a variety of manipulation tasks. These include translational movements along prescribed paths, such as following a laser beam propagation line, or rotational adjustments along the polar and azimuthal angles for component alignment. For placement operations, the system receives target $(x, y)$ coordinates and orientation data. The robot translates the component to the desired location and gently deposits it onto the optical table.

Before finalizing placement, the end-effector camera performs a local verification and correction routine. Minor translational or rotational deviations are adjusted in situ to ensure alignment within the specified tolerance bounds. This dual-stage visual feedback—combining global position estimation with local fine correction—enables highly accurate, repeatable pick-and-place performance, which is critical for the assembly and realignment of complex optical experimental setups. The code for robotic arm control is available in~\cite{AUTO-Arm}.

\subsection{Correction of positional errors}
Despite the high accuracy of the robotic placement system, occasional positional errors were observed. These inaccuracies arise from two primary sources: variable magnetic attraction between component bases and the optical table, and mechanical slippage at the gripper interface due to wear of the rubber contact pads.

To mitigate these errors and ensure reliable and robust assembly of setups, we implemented an automated correction protocol based on post-placement verification. After each placement, the computer vision system re-evaluates the actual position of the optical element and compares it against the target coordinates. If the distance between the detected and intended positions exceeds a predefined tolerance threshold, the robot initiates a re-placement procedure. For all experiments reported in this work, we set this tolerance threshold to 0.5 mm.

This feedback-correction loop ensures that components are repositioned until they fall within the acceptable positional tolerance threshold. Empirically, most components require only a single placement attempt to meet this threshold. While this process may appear inefficient, the average placement time remains significantly lower than that of a human operator. Even with occasional multiple attempts, the robotic system maintains a net speed advantage, as human placement times typically exceed 20 seconds per component.

\subsection{Fine alignment tool}
The fine alignment tool is a compact, robot-deployable device designed to perform precision micro-adjustments on standard optical mounts (Figure~\ref{fig:figure3}b, Supplementary Video 9). The tool integrates an Arduino Nano microcontroller, a custom-designed printed circuit board (PCB) incorporating dual motor controllers, a camera module, and a dedicated illumination system. All components are enclosed in a 3D printed low-profile housing.

The embedded camera is centrally mounted within the housing of the tool and aligned coaxially with the tool’s mechanical axis. A ring of surface-mounted LEDs is positioned directly above the camera, providing uniform illumination and mitigating shadows from external light sources.

The core actuation mechanism consists of two custom-fabricated Allen keys that serve as precision interface tools for engaging standard optical component knobs. These Allen keys are machined from solid stock with a seamless transition between the shaft and tool head, forming a single integrated axle. The design was iteratively refined through multiple fabrication cycles, achieving sub-100 µm tolerances. To minimize mechanical resistance and reduce power consumption, the rotational axes are supported by high-performance bearings and shaft collars, ensuring stable alignment and preventing stalling during actuation.

The tool’s software stack utilizes asynchronous multithreaded control, allowing the Arduino to operate independently of the primary robotic control system. This autonomy enables simultaneous translational and rotational movement, facilitating robust Allen key insertion even in cases of slight misalignment. The insertion routine leverages a helical approach, wherein the tool moves forward while rotating, effectively locating the socket with minimal mechanical stress.

To ensure secure disengagement, the system includes a micro-backdriving feature: upon completing an adjustment, the motors are rotated slightly in the reverse direction of the most recent turn. This subtle motion prevents mechanical binding and ensures smooth retraction of the Allen keys. The combination of precise motor control, low-friction mechanical design, and robust software architecture allows for reliable, repeatable fine adjustments across a variety of optical mounts and experimental configurations.

\subsection{Optical measurements}
For the fabrication of the multilayer photonic crystals (measured in Figure~\ref{fig:figure4}d), we employed plasma-enhanced chemical vapor deposition
(PECVD) using the SAMCO PD-220 system to deposit a stack of silicon (Si) and silica (SiO$_2$) layers onto a fused silica substrate (microscope coverslips of thickness 180 microns) with a deposition rate of approximately 55 nm/min for both materials. The deposition time was used to control the thicknesses of the individual layers. The photonic crystal consists of eight total layers, alternating between Si and SiO$_2$, each with a thickness of 40 nm except the fifth (Si) layer of thickness 110 nm which serves as a defect layer, creating a cavity. The simulated angle- and wavelength-resolved transmission spectrum was calculated using the Rigorous Coupled Wave Analysis (RCWA) method, as implemented in Stanford Stratified Structure Solver (S$^4$)~\cite{StanfordS4}. In the simulations, the material model (n, k) for Si was taken from Ref.\cite{franta2013application} and fit to a fourth order polynomial in the wavelength range of 400-700 nm, while the index of SiO$_2$ was assumed to be 1.45.

\bibliography{references}

\end{document}